\documentstyle[times,pramana,epsfig,floats]{ias}

\begin{document}
\newcommand{\stp}{\tilde t_1}
\def\chio{\textstyle\raise.4ex\hbox{$\textstyle\tilde\chi{^0_1}$}}

\begin{titlepage}

\thispagestyle{empty}
\def\thefootnote{\fnsymbol{footnote}}       

\begin{center}
\mbox{ }

\end{center}
\begin{flushright}
\vspace* {-2.0cm}
\Large
\mbox{\hspace{10.2cm} hep-ph/0609016} \\
\end{flushright}
\begin{center}
\vskip 1.0cm
{\Huge\bf
Small Visible Energy Scalar Top
}
\vspace{2mm}

{\Huge\bf
Iterative Discriminant Analysis
}
\vskip 1cm
{\LARGE\bf A. Sopczak$^1$, A. Finch$^1$, A. Freitas$^2$, \\
          C. Milst\'ene$^3$, M.~Schmitt$^4$
\smallskip

\Large $^1$Lancaster University, UK; $^2$Zurich University, Switzerland; \\
       $^3$Fermilab, USA; $^4$Northwestern University, USA}

\vskip 2.5cm
\centerline{\Large \bf Abstract}
\end{center}

\vskip 2.cm
\hspace*{-0.5cm}
\begin{picture}(0.001,0.001)(0,0)
\put(,0){
\begin{minipage}{\textwidth}
\Large
\renewcommand{\baselinestretch} {1.2}
Light scalar top quarks with a small mass difference with respect to the neutralino mass are
of particular cosmological interest. This study uses an Iterative Discriminant Analysis 
method to optimize the expected selection efficiency at a International Linear Collider (ILC). 
\renewcommand{\baselinestretch} {1.}

\normalsize
\vspace{5.5cm}
\begin{center}
{\sl \large
Presented at the 2006 International Linear Collider Workshop - Bangalore, 
India, \\
to be published in the proceedings.
\vspace{-6cm}
}
\end{center}
\end{minipage}
}
\end{picture}
\vfill

\end{titlepage}

\newpage
\thispagestyle{empty}
\mbox{ }
\newpage
\setcounter{page}{1}

\mark{{Linear Collider Workshop}{A. Sopczak, A. Finch, A. Freitas, C. Milst\'ene, M.~Schmitt}}
\title{\Large Small\,Visible\,Energy\,Scalar\,Top\,Iterative\,Discriminant\,Analysis}

\author{A. Sopczak$^1$, A. Finch$^1$, A. Freitas$^2$, C. Milst\'ene$^3$, M.~Schmitt$^4$}
\address{$^1$Lancaster U., UK, $^2$Zurich U., Switzerland, $^3$Fermilab, USA, $^4$Northwestern U., USA}

\abstract{
Light scalar top quarks with a small mass difference with respect to the neutralino mass are
of particular cosmological interest. This study uses an Iterative Discriminant Analysis 
method to optimize the expected selection efficiency at a International Linear Collider (ILC). 
}

\maketitle
\vspace*{-7mm}
\section{Introduction}
\vspace*{-3.5mm}
The search for scalar top quarks and the determination of their parameters in the framework of
Supersymmetric models are important aspects of the Linear Collider physics programme.
The lightest neutralino with a small mass difference ($\Delta m$) to the scalar top quark
is a candidate for dark matter in the Universe~\cite{Balazs:2004bu}. The dark matter rate
could be measured at a Linear Collider~\cite{carena,sopczak}.
This study applies an Iterative Discriminant Analysis (IDA)~\cite{ida} to a scenario 
involving a 122.5~GeV scalar top and a 107.2~GeV neutralino at $\sqrt{s}=260$~GeV which is near threshold. 
This center-of-mass energy will be used for the scanning phase of the ILC operation and a luminosity of 
50~fb$^{-1}$ is assumed.
The event generation and detector simulation has been performed with unpolarised beams as for a sequential 
cut-based analysis~\cite{carena}. In particular, a vertex detector concept of the Linear Collider Flavour 
Identification (LCFI) collaboration~\cite{lcfi}, which studies pixel detectors for heavy quark flavour 
identification, has been implemented in simulations\,for\,c-quark\,tagging in scalar top\,studies.

\vspace*{-4.5mm}
\section{Preselection}
\vspace*{-3.5mm}
Only two c-quarks and missing energy (from undetected neutralinos) are expected from\,the reaction 
$\rm e^+e^- \rightarrow \stp \bar{\tilde{t}}_1 \rightarrow c \chio \bar c \chio$.
After a simple event preselection, the expected remaining signal and background 
events are given in Table~\ref{tab:pre}.
The requirement $0.1 < E_{\rm vis}/\sqrt{s} < 0.3 $ reduces the
$\rm \rm e^+e^- \to W^+W^-$, ZZ, $\rm q\bar{q}$ and 
$\rm \gamma\gamma \to q\bar{q}$ backgrounds.
Remaining two-photon events are almost completely removed by the cut $p_{\rm t}({\rm event}) > 15$~GeV. 
Requiring at least four but no more than 50 tracks removes mostly 
very low multiplicity background.

\begin{figure}[htbp]
\vspace*{-0.6cm}
\begin{center}
\begin{minipage}{0.49\textwidth}
\epsfig{file=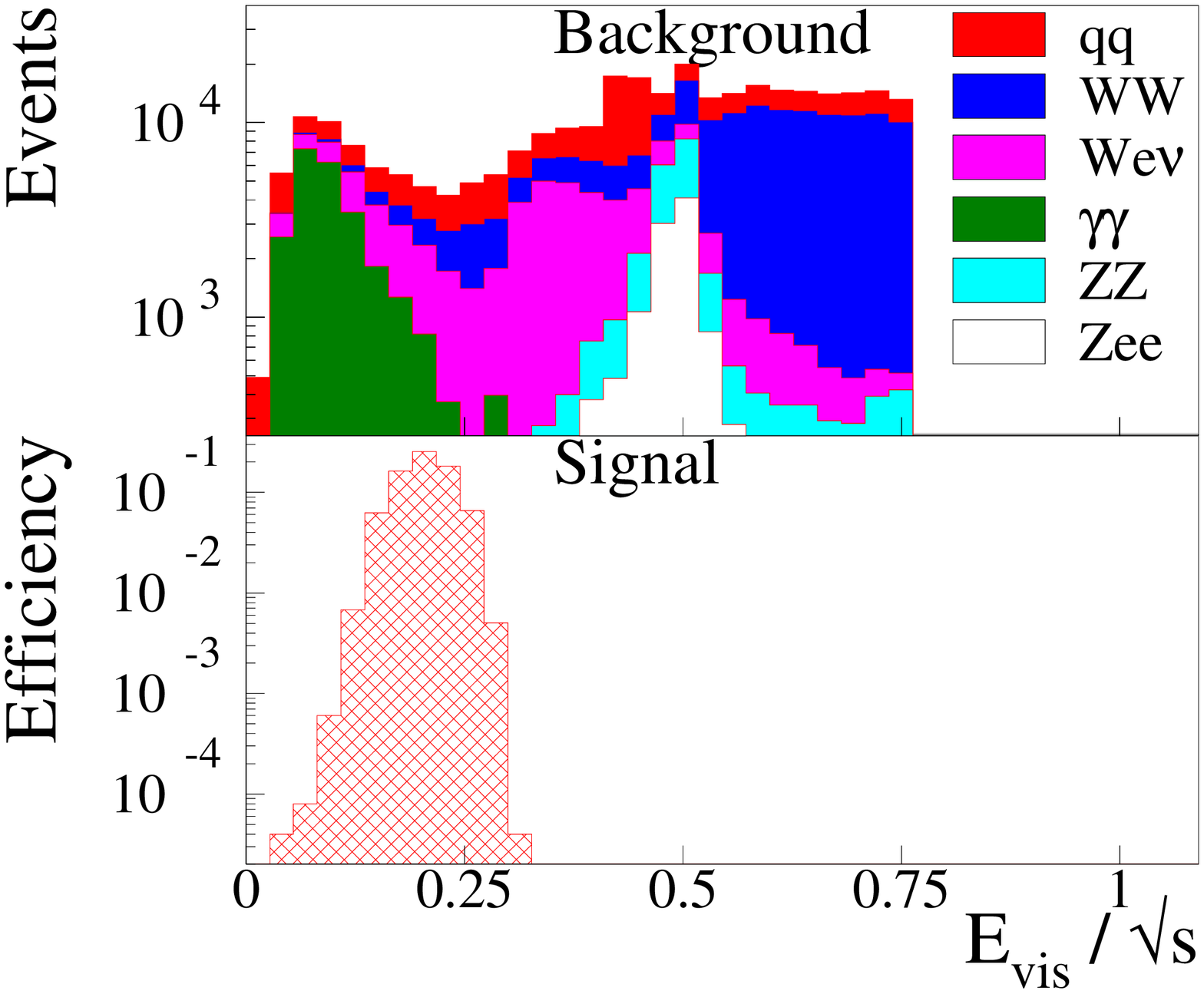,width=\textwidth}
\vspace*{-0.6cm}
\end{minipage}
\begin{minipage}{0.49\textwidth}
\epsfig{file=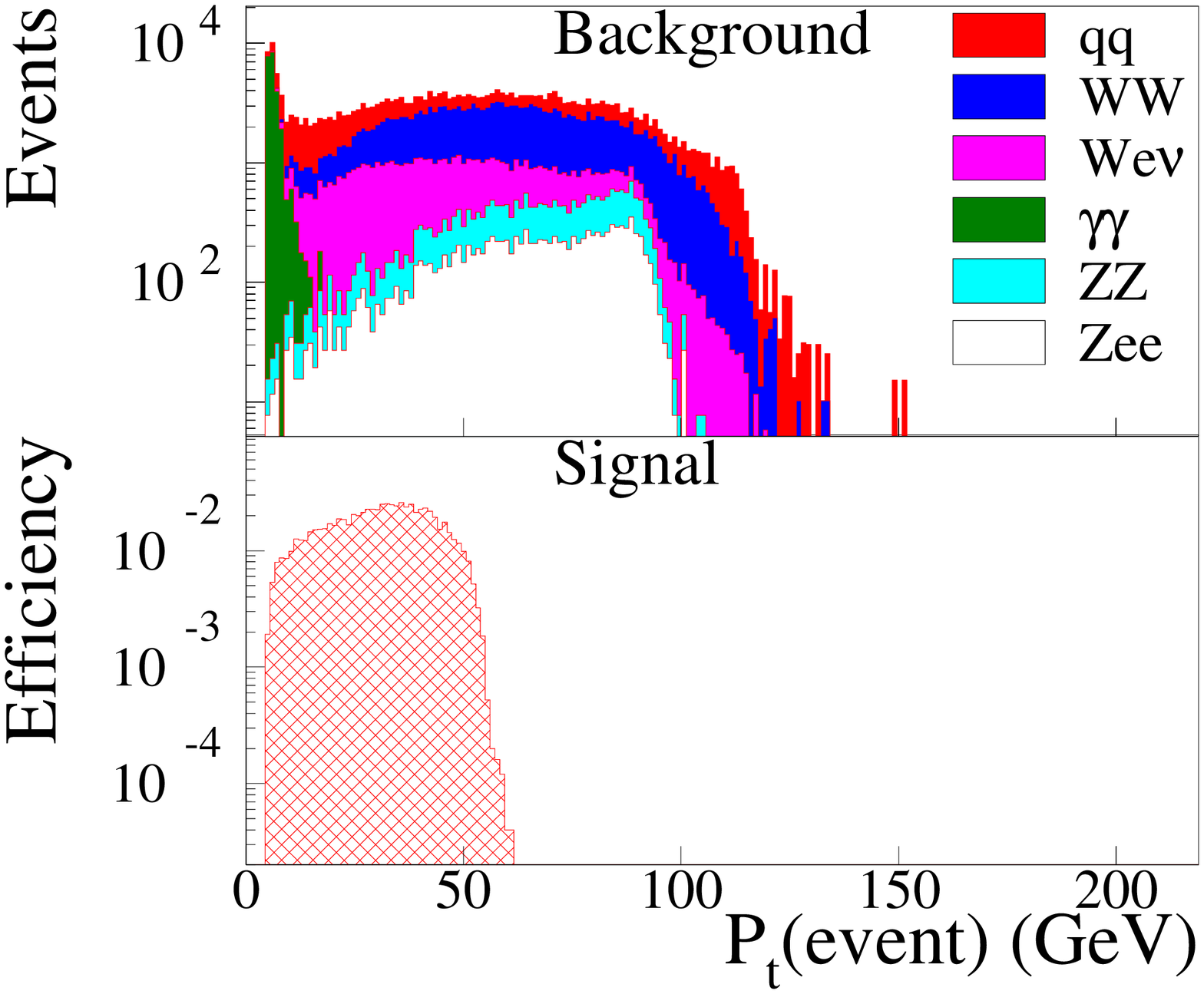,width=\textwidth}
\vspace*{-0.6cm}
\end{minipage}
\begin{minipage}{0.49\textwidth}
\epsfig{file=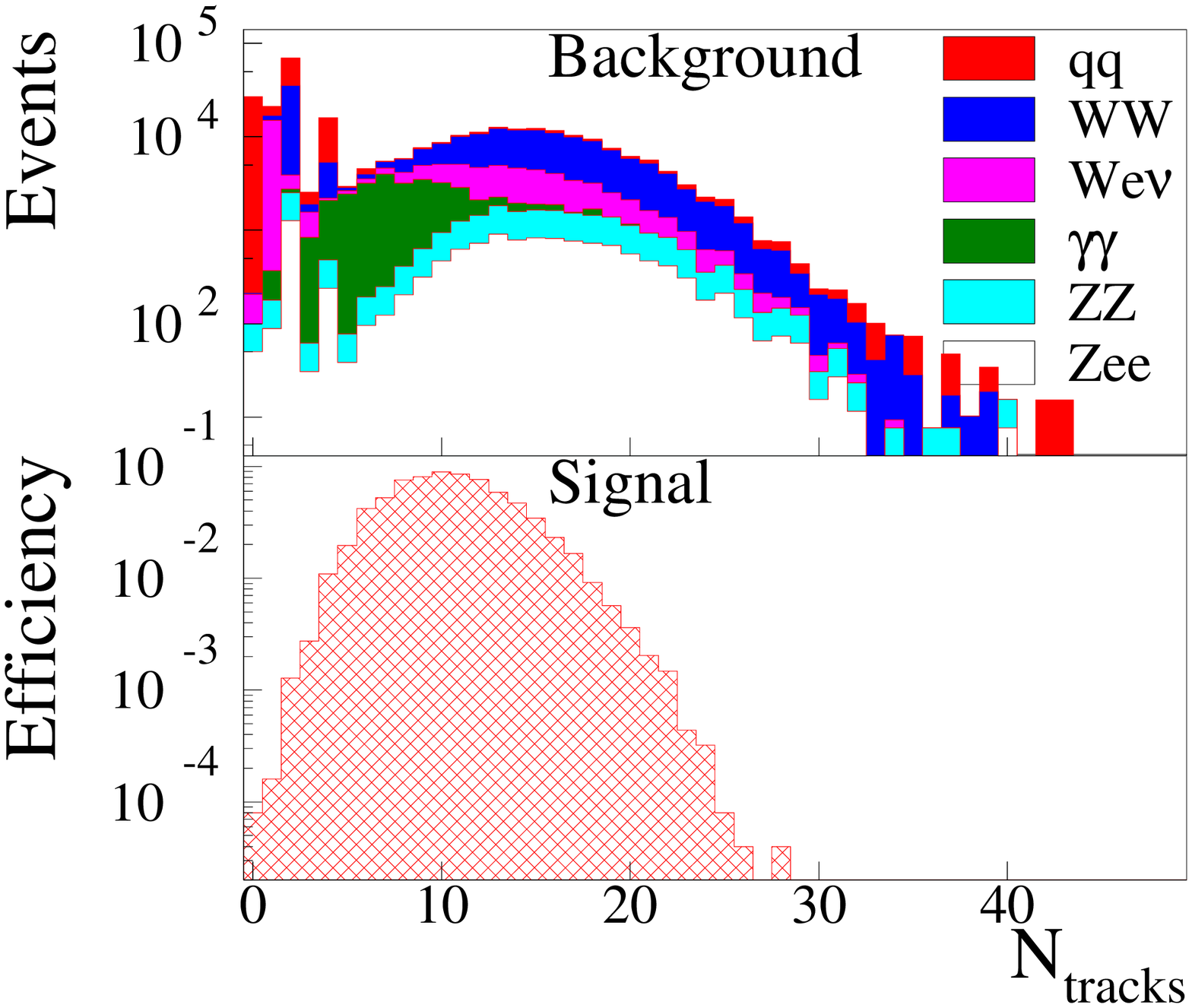,width=\textwidth}
\end{minipage}
\end{center}
\vspace*{-0.6cm}
\caption{Event distributions before preselection.}
\label{fig:pre}
\vspace*{-0.5cm}
\end{figure}

\begin{table}[h]
\vspace*{-1mm}
\caption{Generated events, events used for the IDA training, events after the 
preselection, $\sqrt{s} = 260$ GeV cross section, scaling factor, and expected number of events.}
\begin{tabular}{lrrcccc}
\hline
        & Total         & 50\%     & After  &$\sigma$            & Factor          & Expected     \\
Process & $\times 1000$ & training & preselection  & (pb)         & per 50~fb$^{-1}$& events       \\ 
\hline
signal  & 50     & 25    & 13113                   & 0.0225      & 0.0450          & 590     \\
$\rm q \bar{q}$, $\rm q \neq t$   
         & 350   & 175     &    55                 & 49.5        & 14.14           & 778     \\  
$\rm W^+W^-$ & 180   &  90     &    49             & 16.9        & 9.300          & 456     \\
$\rm W e\nu$ & 210   & 105     &   914             &  1.73       &  0.824          & 753     \\
2-photon & 1600 &  800    &     3                  & 786         & 49.13           & 147     \\        
$\rm ZZ$         & 30    &  15     &    13         &  5.05       & 18.33           & 238     \\
$\rm e e Z$  & 210   & 105     &    12             &  1.12       &  0.533          &   6.4   
\end{tabular}
\label{tab:pre}
\vspace*{-0.7cm}
\end{table}
\clearpage
The distributions of visible energy, transverse momentum and charged track multiplicity 
before preselection are shown in Fig.~\ref{fig:pre}.
After the preselection 52.5\% signal efficiency and 2379 background events remain per 50~fb$^{-1}$.

\section{IDA Event Selection}
\begin{figure}[htbp]
\begin{minipage}{0.49\textwidth}
\epsfig{file=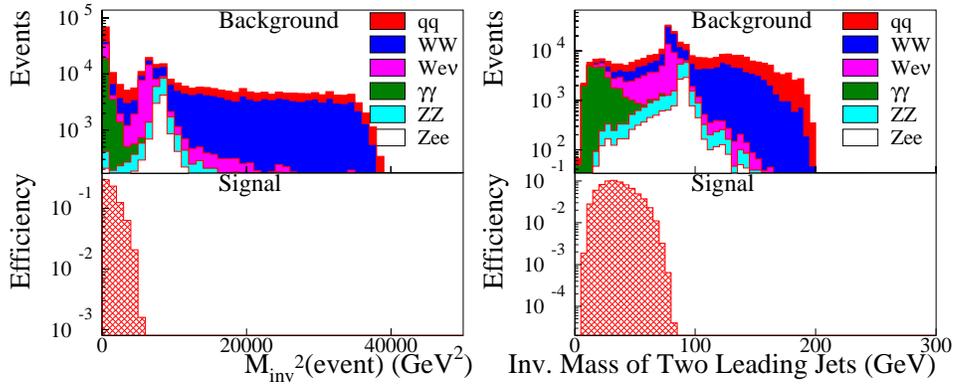,width=\textwidth}
\end{minipage}
\begin{minipage}{0.49\textwidth}
\epsfig{file=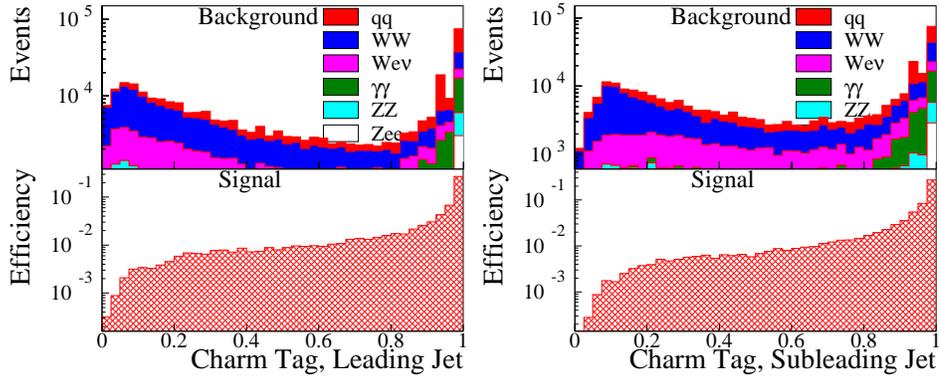,width=\textwidth}
\end{minipage}
\vspace*{-0.5cm}
\caption{Distributions of IDA inputs after preselection.}
\label{fig:mass}
\end{figure}

\begin{figure}[htbp]
\begin{minipage}{0.49\textwidth}
\epsfig{file=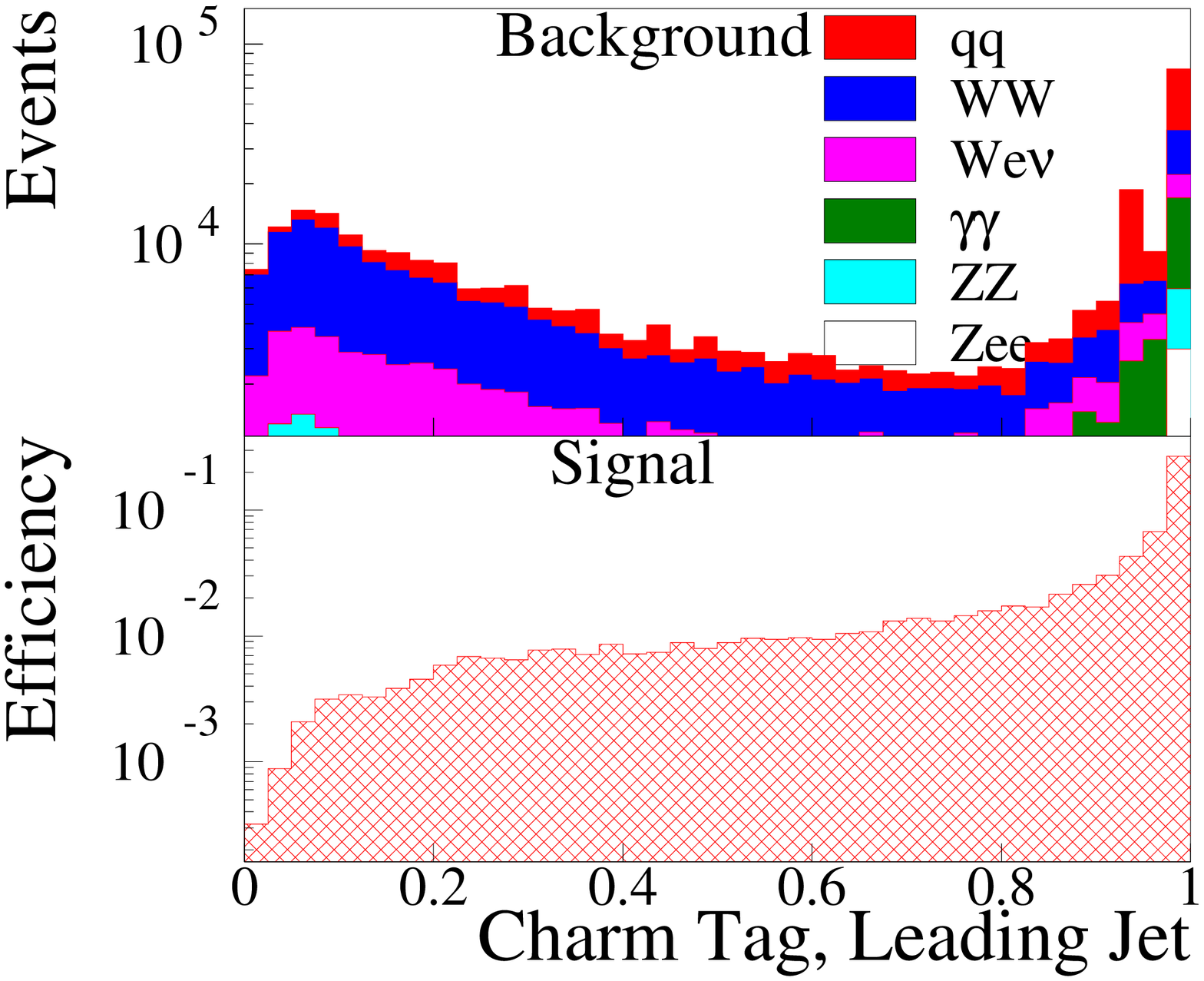,width=\textwidth}
\end{minipage}
\begin{minipage}{0.49\textwidth}
\epsfig{file=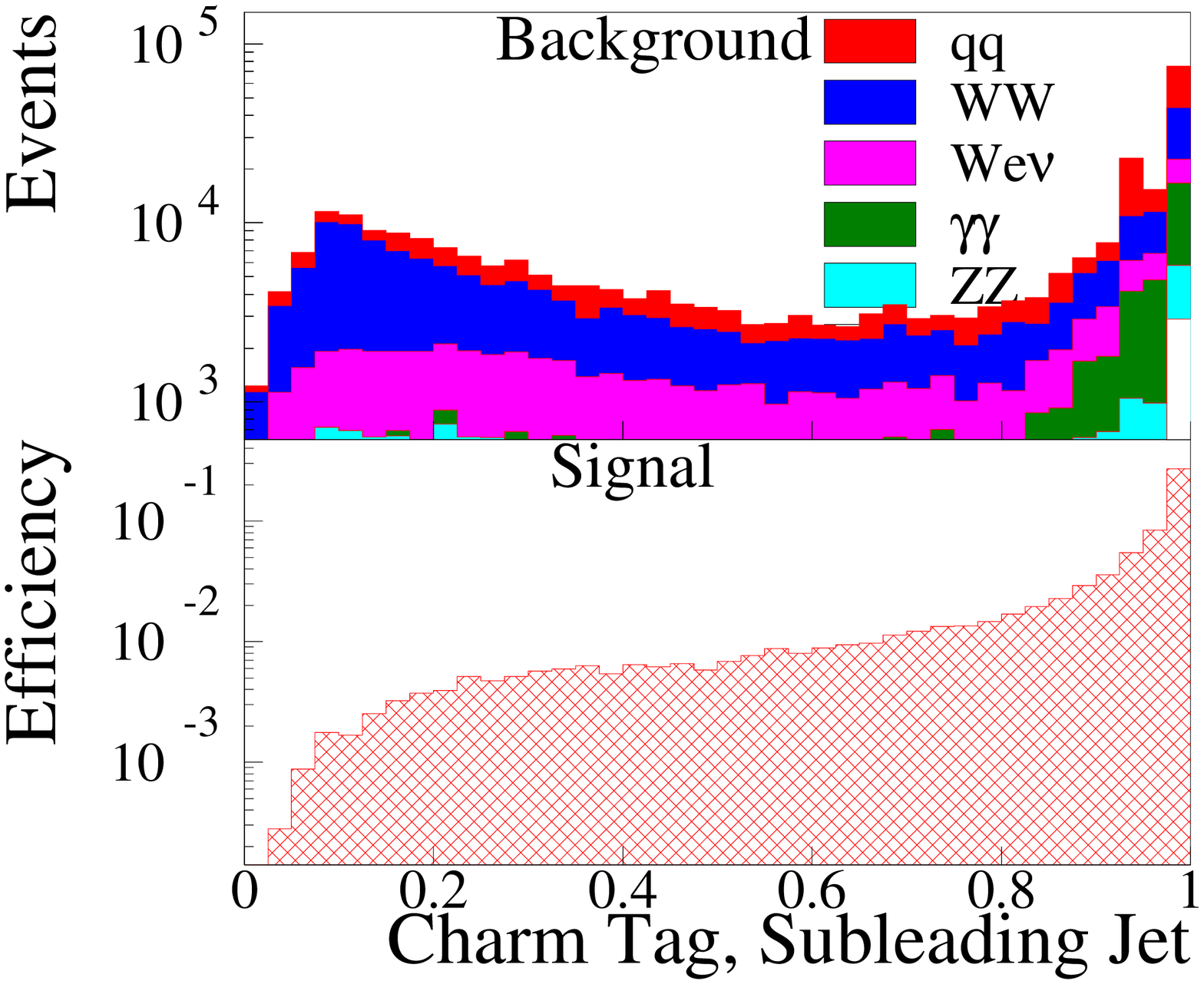,width=\textwidth}
\end{minipage}
\vspace*{-0.5cm}
\caption{Distributions of IDA inputs after preselection.}
\label{fig:ctag}
\end{figure}

\vspace*{-2mm}
Figure~\ref{fig:mass} shows additional input variables used in the IDA:
the event invariant mass and the invariant mass of the two jets. 
Further input variables are the c-quark tagging of the leading (most energetic) 
and subleading jets (Fig.~\ref{fig:ctag}). The c-quark jet tagging has been performed
with a neutral network~\cite{kuhl} optimized for small $\Delta m$.

\begin{figure}[htbp]
\begin{minipage}{0.49\textwidth}
\epsfig{file=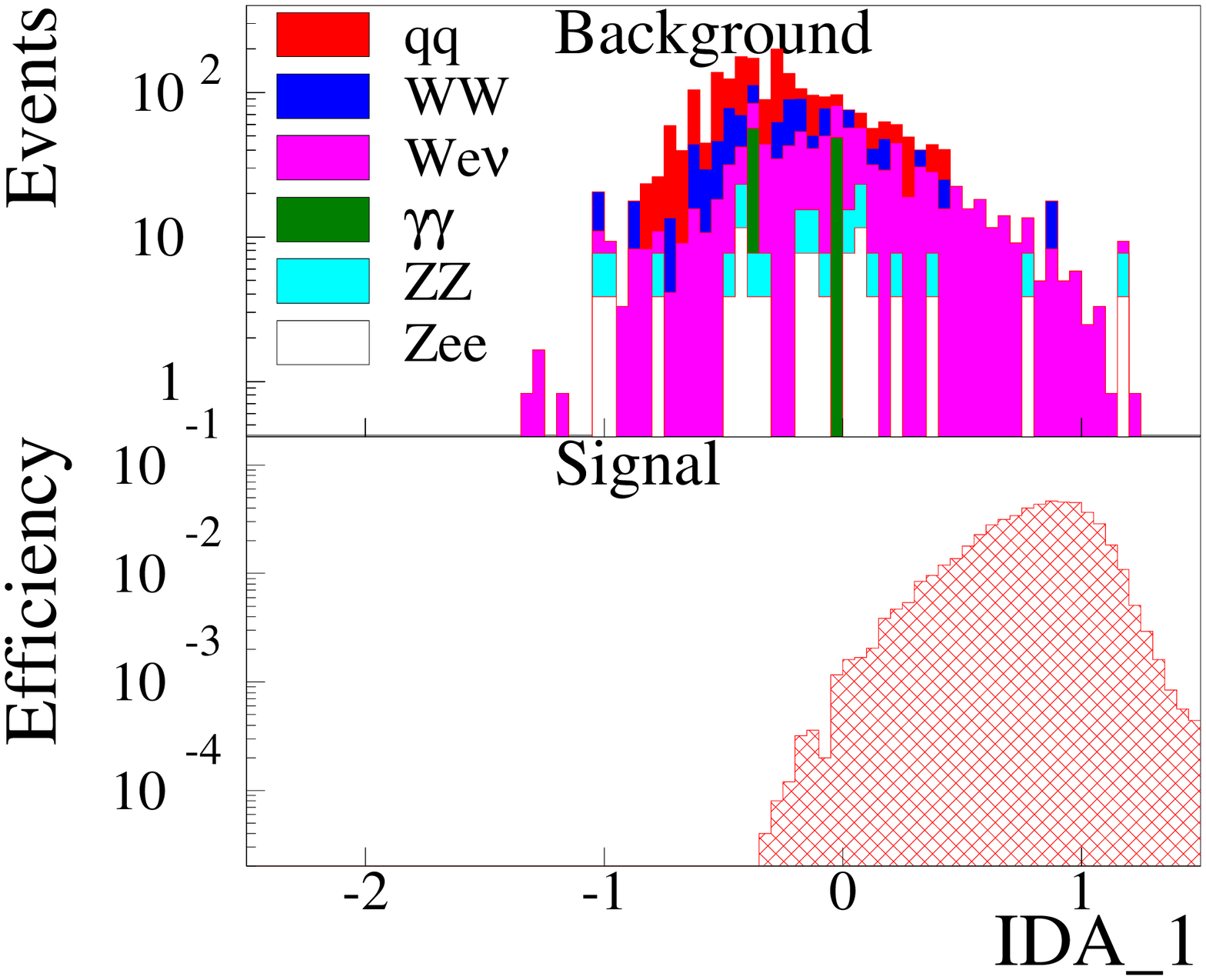,width=\textwidth}
\end{minipage}
\begin{minipage}{0.49\textwidth}
\epsfig{file=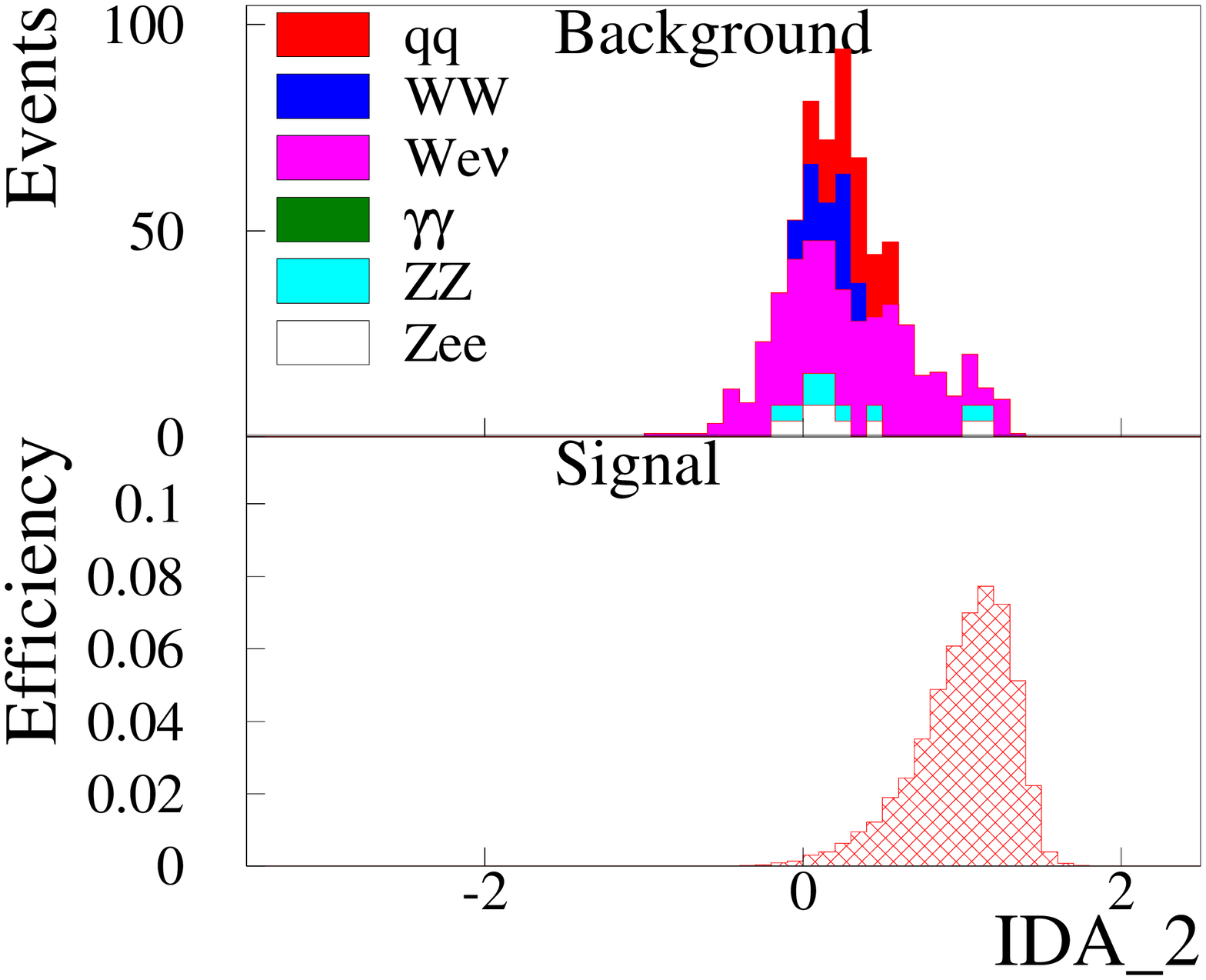,width=\textwidth}
\end{minipage}
\vspace*{-0.5cm}
\caption{IDA output for steps 1 and 2.
         In the first IDA step, a cut on IDA\_1 is applied at zero,
         retaining 99.5\% of the simulated signal input events.}
\label{fig:ida}
\end{figure}

\begin{figure}[htbp]
\vspace*{-0.8cm}
\begin{center}
\epsfig{file=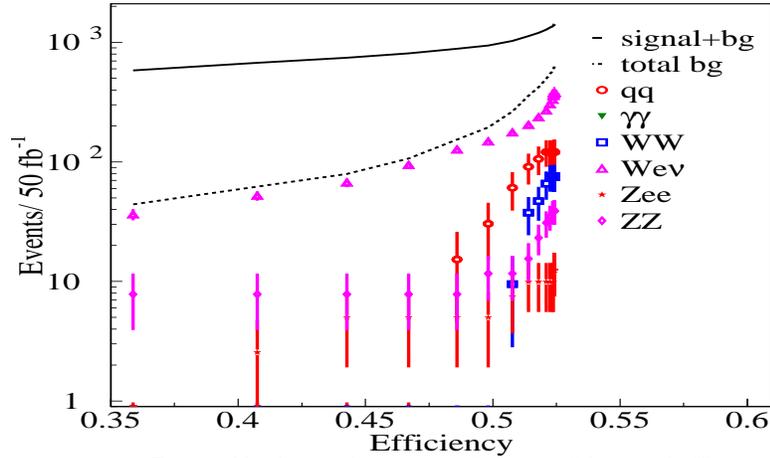,width=0.8\textwidth,height=0.36\textheight}
\end{center}
\vspace*{-0.5cm}
\caption{Expected background events as a function of the signal efficiency.}
\label{fig:perf}
\vspace*{-0.8cm}
\end{figure}

The IDA has been applied in two steps in order to optimize the performance, as shown in Fig.~\ref{fig:ida}. 
In the first step, a cut was applied on the IDA\_1 output variable such that 99.5\% of the 
signal events remain.
This leads to 52.0\% signal efficiency and 490 background events per 50 fb$^{-1}$.
These remaining signal and background events have been passed to the second IDA step. 
A cut on the IDA\_2 output variable determines the final selection efficiency and the 
corresponding expected background.
The resulting performance is shown in Fig.~\ref{fig:perf}.

\clearpage

\section{Conclusions}
\vspace*{-2mm}
The Iterative Discriminant Analysis leads to a good signal over background ratio
for the investigated Supersymmetric scenario with a scalar top mass of 122.5~GeV and a neutralino
mass of 107.2~GeV (Fig.~\ref{fig:perf}). 
For 50\% signal efficiency (560 signal events) about 200 background events (mostly $\rm We\nu$) 
are expected per 50~fb$^{-1}$.
The analysis is a step towards a precise scalar top mass determination. 
The expected uncertainty in the light scalar top mass measurement dominates the 
uncertainty in the dark matter prediction from the co-annihilation process~\cite{carena,sopczak}. 
A further study will also include a systematic error analysis.
While this study has focused on $\sqrt{s}=260$~GeV, an IDA study for this small $\Delta m$ scenario 
at $\sqrt s = 500$~GeV is in progress~\cite{susy06} and
a much improved determination of the dark matter prediction is expected~\cite{ayres}.
Further plans are to focus on the vertex detector design, including the implementation of a new 
LCFI~\cite{lcfi} c-quark tagging.

\vspace*{-4mm}
\section*{Acknowledgements}
\vspace*{-2mm}
AS would like to thank the LCFI Colleagues for fruitful discussions on the analysis and for comments 
during the preparation of the presentation.

\end{document}